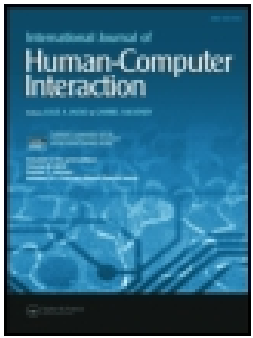

# International Journal of Human–Computer Interaction

# A Systematic Literature Review of User Trust in AI-Enabled Systems: An HCI Perspective

Tita Alissa Bach, Amna Khan, Harry Hallock, Gabriela Beltrão & Sonia Sousa

Published online: 10 Nov 2022.

Submit your article to this journal ⬈

Article views: 1330

View related articles ⬈

View Crossmark data ⬈

Taylor & Francis
Taylor & Francis Group

# A Systematic Literature Review of User Trust in AI-Enabled Systems: An HCI Perspective

Tita Alissa Bach[a]*, Amna Khan[b]*, Harry Hallock[a], Gabriela Beltrão[b], and Sonia Sousa[b]

[a]Group Research and Development, DNV AS, Høvik, Norway; [b]School of Digital Technologies, Tallinn University, Tallinn, Estonia

**ABSTRACT**

User trust in Artificial Intelligence (AI) enabled systems has been increasingly recognized and proven as a key element to fostering adoption. It has been suggested that AI-enabled systems must go beyond technical-centric approaches and towards embracing a more human-centric approach, a core principle of the human-computer interaction (HCI) field. This review aims to provide an overview of the user trust definitions, influencing factors, and measurement methods from 23 empirical studies to gather insight for future technical and design strategies, research, and initiatives to calibrate the user-AI relationship. The findings confirm that there is more than one way to define trust. Selecting the most appropriate trust definition to depict user trust in a specific context should be the focus instead of comparing definitions. User trust in AI-enabled systems is found to be influenced by three main themes, namely socio-ethical considerations, technical and design features, and user characteristics. User characteristics dominate the findings, reinforcing the importance of user involvement from development through to monitoring of AI-enabled systems. Different contexts and various characteristics of both the users and the systems are also found to influence user trust, highlighting the importance of selecting and tailoring features of the system according to the targeted user group's characteristics. Importantly, socio-ethical considerations can pave the way in making sure that the environment where user-AI interactions happen is sufficiently conducive to establish and maintain a trusted relationship. In measuring user trust, surveys are found to be the most common method followed by interviews and focus groups. In conclusion, user trust needs to be addressed directly in every context where AI-enabled systems are being used or discussed. In addition, calibrating the user-AI relationship requires finding the optimal balance that works for not only the user but also the system.

## 1. Introduction

Various real-world applications of Artificial Intelligence (AI) have been developed and implemented to improve, for example, online health platforms (Panda & Mohapatra, 2021), banking systems (Mohapatra, 2021), businesses, industry (Mohapatra & Kumar, 2019), and life in general (Abebe & Goldner, 2018; Banerjee et al., 2021, 2022; Davenport & Ronanki, 2018). Nevertheless, with this uptake, more concerns have been raised about certain AI characteristics, such as being opaque ("black box"), uninterpretable, and biased, and the risks these characteristics impose (Gulati et al., 2019; Rai, 2020). AI opaqueness, for example, makes it harder to predict how AI may behave (Bathaee, 2017; Fainman, 2019), may make back-tracking errors and decisions more difficult (Bathaee, 2017), and the effort to understand the logic of how an output is produced harder (Fainman, 2019). These difficulties are amplified with the fact that AI output is suggested to be inherently uncertain (Zhang & Hußmann, 2021). When poorly designed or adapted to target users, AI usage could mislead users into unfair and even incorrect decision-making

(Lakkaraju & Bastani, 2020). Consequently, the real-world consequences of a failed AI-enabled system can be catastrophic, leading to, for example, discrimination (Buolamwini, 2017; Buolamwini & Gebru, 2018; Dastin, 2022; Hoffman & Podgurski, 2022; Kayser-Bril, 2020; Olteanu et al., 2019; Ruiz, 2019), and even death (Kohli & Chadha, 2020; Pietsch, 2021). Here, AI-enabled systems are defined as AI systems with capabilities to improve existing systems' performance, i.e., AI-enhanced systems (Boland & Lyytinen, 2017), for example, recommender systems, and/or AI systems with capabilities to develop new applications, i.e., AI-based systems (Wuenderlich & Paluch, 2017), for example, virtual agents and robotic surgery (Rzepka & Berger, 2018).

The potential negative consequences of using AI-enabled systems have led to a lack of trust by users, and highlighted the importance of ethics. As Bryson (2019) stresses, AI-enabled systems are an artifact designed by humans who supposedly should be held accountable for outcomes regardless whether humans understand the logic behind a particular process followed by the systems. As a result, efforts have been put into the creation of AI ethics guidelines to address this issue (Jobin et al., 2019). Nevertheless, a review of 84 AI ethics guidelines in different countries reveals that although there are similarities in the principles proposed by these guidelines (i.e., transparency, justice and fairness, non-maleficence, responsibility and privacy) (Jobin et al., 2019), there are differences in the interpretation, prioritization, and implementation of the ethical principles. Consequently, AI ethics guidelines can in fact be misplaced and harmful if the efforts distract the focus to operationalize ethical AI-enabled systems (Mittelstadt, 2019; Munn, 2022). Efforts thus have been broadened to the concept of trustworthy AI that includes AI-enabled systems that are not only ethical, but also lawful and robust (European Commission, 2019).

Trustworthiness in AI-enabled systems can be achieved by making sure that the risks associated with certain characteristics of AI-enabled systems are managed (Cheatham et al., 2019; Floridi et al., 2018). Accordingly, AI developers, researchers and regulators suggest that AI-enabled systems must go beyond technical-centric approaches and towards embracing a more human-centric approach (Shneiderman, 2020a), a core principle of the human-computer interaction (HCI) field (Xu, 2019; Zhang & Hußmann, 2021). According to Hoffman et al. (2001), HCI is a multidisciplinary field of study that focuses on the development, evaluation, and dissemination of technology to meet users' needs by optimizing how users and technology interact. HCI has broadened its focus during the third wave of computing when technology was embedded into practically all sectors and computers were connected by the internet. Today, HCI covers computer science, engineering, cognitive science, ergonomics, design principles, economics, and behavioral and social sciences to meet rapidly changing user needs (Rogers, 2012). Importantly, HCI has been used to ensure trustworthiness of AI in recent years in the form of, for example, guidelines, frameworks, and principles (Leijnen et al., 2020; Robert et al., 2020; Shneiderman, 2020a; Smith, 2019). This is because utilizing an HCI approach assumes an interdisciplinary view of technology (Rogers, 2012), and thus relies on knowledge from, among others, psychology, sociology, and computer science to develop strategies to foster user trust in AI-enabled systems (Corritore et al., 2007; Robert et al., 2020).

## 1.1. Research scope

A growing number of researchers argue that fostering and maintaining user trust is the key to calibrating the user-AI relationship (Jacovi et al., 2021; Shin, 2021), achieving trustworthy AI (Smith, 2019), and further unlocking the potential of AI for society (Bughin et al., 2018; Cheatham et al., 2019; Floridi et al., 2018; Taddeo & Floridi, 2018). Despite trust being a concept widely studied across many disciplines (Mcknight & Chervany, 1996), the definition, importance and measurement of user trust in AI-enabled systems are not yet as well agreed on and studied (Bauer, 2019; Sousa et al., 2016). Consequently, the terms trust, trustworthy, and trustworthiness may be addressed without a clear focus and understanding of how these concepts may play a role in the interactions between users and AI-enabled systems, as principles alone cannot guarantee actual trustworthiness (Mittelstadt, 2019). Trust is not a direct measure of value and cannot be framed in a single construct. Instead, trust in technology is a social-technical construct and reflects an individual's willingness to be vulnerable to the actions of another, irrespective of the ability to monitor or control these actions (Mayer et al., 2006). In addition, AI-enabled systems are inherently "complex" (Bathaee, 2017; Fainman, 2019; Mittelstadt, 2019), meaning that their functions and design elements make it more challenging for users to immediately understand, accept and justify (Bathaee, 2017; Fainman, 2019; Mittelstadt, 2019). Even when users feel they can control a complex system, they are known to misinterpret the causality of the elements within it (Dörner, 1978). These complex characteristics can pose a challenge to address user trust in AI-enabled systems and risk a trust gap between users and the systems (Ashoori & Weisz, 2019). The current review focuses on the user-AI relationship because AI is likely to amplify the importance of collaboration between the user and the system, in which the system is one of the collaborative partners in the relationship.

To our knowledge, there is still little research focused on providing an overview of empirical studies focused on user trust in AI-enabled systems where the user-AI relationship is the center point (Xu, 2019). Therefore, our study aims to contribute to the HCI literature by providing an overview of user trust definitions, user trust influencing factors, and methods to measure user trust in AI-enabled systems. HCI is naturally used as a search term in this systematic literature review to specifically identify studies that draw on HCI concepts whilst focusing on the user-AI relationship. Findings can be used to provide insight for future technical and design strategies as well as research and initiatives focused on fostering and maintaining user trust in AI-enabled systems.

## 1.2. Research questions

This systematic literature review aimed to answer: how is user trust in AI-enabled systems defined (RQ1)? What factors influence user trust in AI-enabled systems (RQ2)? How can user trust in AI-enabled systems be measured (RQ3)?

## 2. Methodology

### 2.1. Literature search and strategy

Our initial quick search on the topic revealed different approaches to studying user trust in AI-enabled systems. In addition to allowing others to replicate our study, a systematic literature review was chosen because this method provides an overview of the available empirical evidence by reviewing the relevant literature on the chosen topic rigorously and systematically with the aim to minimize bias and produce more reliable results (*About Cochrane Reviews*, n.d.;

**Table 1.** The search terms.

| |
|---|
| (Trustworthy AI OR Trustworthy AI Technology OR Measure Trustworthiness OR Design Trustworthiness OR Explainable AI OR Trust Frameworks) |
| AND |
| (Artificial Intelligence OR Machine Learning OR Predictive Model OR Deep Learning OR Neural Network OR Interpretable Machine Learning OR AI Technology OR Explainable Machine Learning) |
| AND |
| (Human Computer Interaction OR hci OR Human-Computer Interaction OR Human-Computer Trust) |
| AND |
| (Trust in Technology OR AI Technologies OR Trustworthiness) |
| AND |
| (Measure Trust OR Empirical Research OR Trusted Interactions OR Methods OR Tools) |

**Table 2.** Results of the search using EBSCO Discovery Service and Web of Science.

| Database search results | EBSCO Discovery Service | Web of Science |
|---|---|---|
| All results | 641 articles | 16 articles |
| Peer Reviewed | 523 article | 16 articles |
| Language: English | 522 article | 16 articles |
| Publication duration: [2011–2021] | 478 articles | 16 articles |
| Scientific articles and conference proceedings | 478 articles | 15 articles |

Tanveer, n.d.). This method has also been used to, for example, build a theoretical framework for the adoption of AI-enabled systems (Banerjee et al., 2021, 2022; Panda & Mohapatra, 2021).

The systematic literature review was conducted in line with the PRISMA standards for qualitative synthesis (Moher et al., 2010). For this purpose, two computer science digital libraries were used: (1) the "EBSCO Discovery Service" and (2) "Web of Science". The keywords and search strings were selected based on the research questions and a pilot search to ensure all the relevant articles were included. ACM computing classification system (CSS) 2012 was consulted to further refine the search terms strategy (Rous, 2012), resulting in the final keywords chosen as shown in Table 1.

### 2.2. The study selection

Inclusion and exclusion criteria were developed to define the scope of the study as followed: (1) published in English between January 1, 2011 and May 15, 2021, (2) a scientific article or conference proceeding, (3) empirical, (4) presented factors influencing user trust in AI-enabled systems, (5) presented a clear methodology, and (6) a full-text article version available. The search resulted in a total of 493 articles (Table 2). A series of virtual meetings were held to review and discuss the 67 articles included in the full-text screening. Group consensus was used throughout the article selection process to resolve any disagreements regarding eligibility. The first authors (AK, TAB) conducted an additional quality check independently for eligibility. All authors accepted the final 23 articles to be included in the analysis and synthesis (Figure 1).

### 2.3. Analysis and synthesis

Analysis and synthesis were first conducted on three randomly selected articles of the final 23 articles to evaluate the inclusion and exclusion criteria and determine the synthesis process. Authors then independently extracted the following information from the 23 included articles: the title and citation, author(s), the year of publication, the geographical location of the data collection/study, the study focus, the study area/domain/industry, the AI-enabled system being studied, trust definition, methodology, the method to measure user trust, number of participants/dataset, and factors influencing user trust. A series of meetings were held to discuss the process and results of the analyses. Once all authors agreed on the results, the first-authors (AK, TAB) quality checked extracted information from each article for consistency. Common themes of factors influencing user trust were clustered together using content analysis. Group consensus was used throughout the analysis and to resolve any disagreements.

## 3. Results

Approximately half (52.17%) of the 23 included articles were published after 2018 (Table 3). Slightly more than half (56.52%) conducted their studies in the USA and Germany, and 52.17% of the studies were focused on Robotics and E-commerce. The included articles cover various types of AI-enabled systems in which the most common types were general AI/ML and automated algorithms (30.43%). Almost 78.36% articles focused on assessing, predicting, or augmenting antecedents, predictors, critical dimensions and factors of user trust in AI-enabled systems. Moreover, surveys were found to be the most common method to measure user trust (59.56%), either as a standalone or in combination with interviews or focus groups.

### 3.1. RQ1: How is user trust in AI-enabled systems defined?

Seven articles provided trust definitions (Table 4). Eight articles conceptualized trust, but did not define it (Corritore et al., 2012; Duffy, 2017; Elkins & Derrick, 2013; Höddinghaus et al., 2021; Law et al., 2021; Lee et al., 2021; Sharma, 2015; Smith, 2016), and the remaining eight articles neither defined nor conceptualized trust. Four articles used Mayer's trust definition (Mayer et al., 2006) (Foehr & Germelmann, 2020; Glikson & Woolley, 2020; Lin et al., 2019; Thielsch et al., 2018). Two articles (Hoffmann &

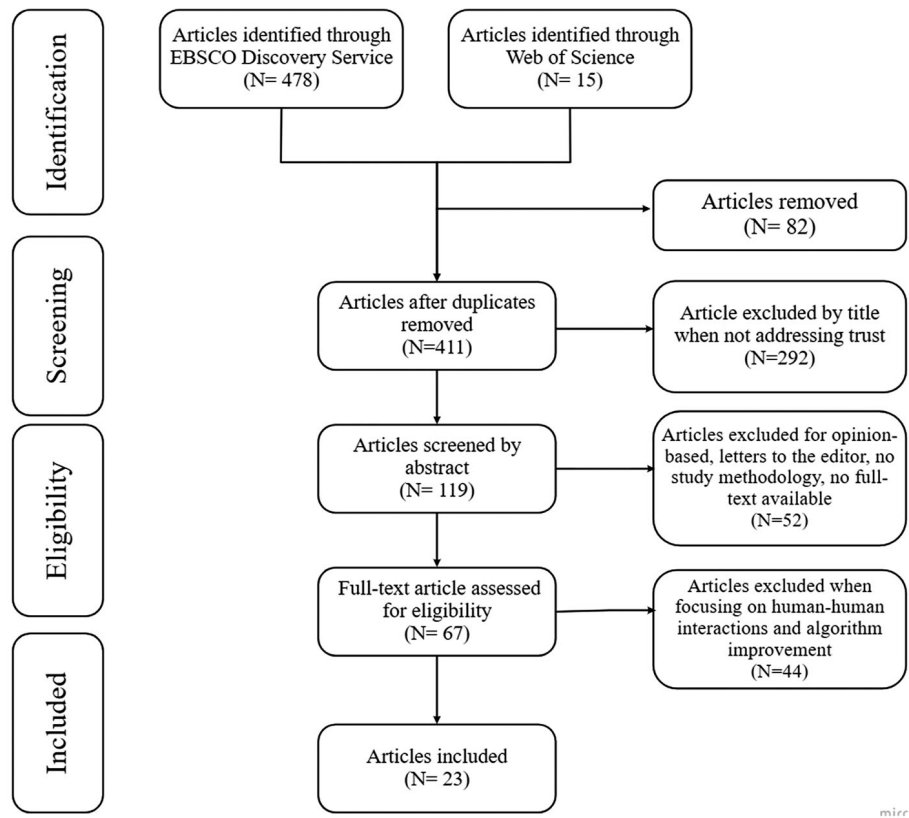

**Figure 1.** PRISMA flow chart of the study selection process.

Söllner, 2014; Zhou et al., 2020) used Lee and See's trust definition (2004). One article (Yan et al., 2013) developed its own definition in combination with citing trustworthy characteristics from Avizienis et al. (2004).

### 3.2. RQ2: What factors influence user trust in AI-enabled systems?

Three main themes were identified from the 23 included articles: socio-ethical considerations, technical and design features, and user characteristics (Table 5). Eight articles identified socio-ethical considerations, 12 articles identified technical and design features, and 22 articles identified user characteristics.

#### 3.2.1. Socio-ethical considerations influencing user trust

Preparing and adjusting the environment where an AI-enabled system was (to be) implemented were suggested as crucial to ensure initial user trust (Lee et al., 2021). This was because the development of AI-enabled systems was often faster than the readiness of their potential users, and a mismatch of readiness levels might lead to low user trust. It was suggested to set up mechanisms in place to foster, maintain, and recover user trust (Binmad et al., 2017), by, for example, ensuring user data protection (Foehr & Germelmann, 2020), encouraging high-quality user interactions (Lin et al., 2019), and solution-oriented technical support (Thielsch et al., 2018). It was also suggested that user trust was likely to increase over time (Elkins & Derrick,

2013). Therefore, building and maintaining open communication with users, for example, by requesting ongoing feedback of an AI-enabled system being used, was suggested as a determinant for user trust (Elkins & Derrick, 2013).

Setting up ethical-legal boundaries of AI-enabled systems was consistently seen as a significant challenge due to unclear accountability between involved parties and unclarity of a determinant if harm occurred to the users (O'Sullivan et al., 2019). Accountability for any harm to the user was put on the manufacturer if a manufacturing defect was identified, the operator if the use of the system was implicated, or the person responsible for performing the maintenance or adjustment if the system failure was rooted in its maintenance or adjustments (O'Sullivan et al., 2019). However, in practice, this pragmatic imputation might not be as simple, for example, in cases where the manufacturer no longer existed, or where the root-cause of damage was unclear (O'Sullivan et al., 2019).

#### 3.2.2. Technical and design features influencing user trust

In developing a virtual agent with the purpose to assist and communicate with a user (e.g., chatbots, embodied conversational agents, smart speakers), the following technical and/or design features were found to increase user trust: (1) anthropomorphism and human-like features, especially benevolent features (e.g., smiling, showing interest in the user) in an AI-enabled system (Elkins & Derrick, 2013; Foehr & Germelmann, 2020; Law et al., 2021; Morana et al., 2020), (2) immediacy behaviours in which the AI-enabled system could create and project a perception of physical and psychological closeness to the user (Glikson & Woolley, 2020), (3) social presence of the AI-enabled system (Glikson & Woolley, 2020; Morana et al., 2020; Weitz et al., 2021), (4) integrity of the AI-enabled system (i.e., repeatedly satisfactory task fulfillment) (Foehr & Germelmann, 2020; Höddinghaus et al., 2021), (5) additional text/speech output when communicating with users (Weitz et al., 2021), (6) providing users with texts rather than a synthetic voice (Law et al., 2021), and (7) a lower vocal pitch of the AI-enabled system (Elkins & Derrick, 2013).

**Table 3.** Overview of the 23 included articles.

| Overview | | N | % |
|---|---|---|---|
| The year of publication | 2012 | 1 | 4.35 |
| | 2013 | 2 | 8.70 |
| | 2014 | 1 | 4.35 |
| | 2015 | 1 | 4.35 |
| | 2016 | 2 | 8.70 |
| | 2017 | 2 | 8.70 |
| | 2018 | 2 | 8.70 |
| | 2019 | 5 | 21.74 |
| | 2020 | 5 | 21.74 |
| | 2021 | 2 | 8.70 |
| | **TOTAL** | **23** | **100** |
| The geographical location of the data collection/study | USA | 7 | 30.43 |
| | Germany | 6 | 26.09 |
| | Multiple countries | 5 | 21.74 |
| | Australia | 1 | 4.35 |
| | China | 1 | 4.35 |
| | India | 1 | 4.35 |
| | Netherlands | 1 | 4.35 |
| | No information | 1 | 4.35 |
| | **TOTAL** | **23** | **100** |
| The study area/domain/industry | Robotics | 6 | 26.09 |
| | E-commerce | 6 | 26.09 |
| | Healthcare | 4 | 17.39 |
| | Business | 2 | 8.70 |
| | Application software | 2 | 8.70 |
| | Finance | 1 | 4.35 |
| | Logistics | 1 | 4.35 |
| | Research | 1 | 4.35 |
| | **TOTAL** | **23** | **100** |
| Types of AI-enabled systems | General AI/ML and automated algorithms | 7 | 30.43 |
| | Virtual agents (e.g., chatbots, embodied conversational agents, smart speakers) | 6 | 26.09 |
| | Websites (e.g., online search, recommender systems) | 6 | 26.09 |
| | Software applications (e.g., mobile applications) | 2 | 8.70 |
| | Complex information systems | 1 | 4.35 |
| | Tracking devices | 1 | 4.35 |
| | **TOTAL** | **23** | **100** |
| Study focus | Assessing user trust | 18 | 78.36 |
| | Testing or applying user trust models or theories | 3 | 13.04 |
| | Developing ethics, regulation and legal framework for user trust | 1 | 4.35 |
| | Preventing artificial divide for user trust | 1 | 4.35 |
| | **TOTAL** | **23** | **100** |
| Methods to measure user trust | Survey | 13 | 56.52 |
| | Survey and interviews/focus group | 3 | 13.04 |
| | Literature review | 3 | 13.04 |
| | Multiple qualitative methods | 3 | 13.04 |
| | Simulation experiment | 1 | 4.35 |
| | **TOTAL** | **23** | **100** |

Specifically for AI/ML and automated algorithms, the following technical and/or design features were found to influence user trust: (1) explanations and information regarding: how the algorithm worked (Glikson & Woolley, 2020), AI's actions (Barda et al., 2020; Glikson & Woolley, 2020; O'Sullivan et al., 2019), reflections of AI reliability (Barda et al., 2020), model performance (Zhang, Genc, et al., 2021), feature influence methods, risk factors to predictive models

(Barda et al., 2020), contextual information (Barda et al., 2020), and interactive risk explanation tools (baseline risk and risk trends) (Barda et al., 2020), (2) correctness of AI/ML predictions (Zhang, Genc, et al., 2021), and (3) AI/ML integrity (Höddinghaus et al., 2021).

One article, which focused on user trust in complex information systems, found that system reliability (dependability, lack and correctness of data, technical verification, distribution of the system) and the quality of the system information (credibility) influenced user trust (Thielsch et al., 2018). Importantly, users with high dependency on the systems and users who had to use the systems, had no other choice but to trust the systems (Thielsch et al., 2018). When an information system used a website to interact with users, multimedia features, security certificate/logo, contact information, and a social networking logo were found to be important for user trust (Sharma, 2015).

**Table 4.** Overview of trust definitions from seven articles.

| Title | Trust definition | Trust definition references |
|---|---|---|
| Alexa, can I trust you? Exploring consumer paths to trust in smart voice-interaction technologies (Foehr & Germelmann, 2020) | The willingness of a party to be vulnerable to the actions of another party based on the expectation that the other will perform a particular action important to the trustor, irrespective of the ability to monitor or control that other party. | (Mayer et al., 2006) |
| Building e-commerce satisfaction and boosting sales: the role of social commerce trust and its antecedents (Lin et al., 2019) | The willingness of a party [the trustor] to be vulnerable to the actions of another party based on the expectation that the other [the trustee] will perform a particular action important to the trustor, irrespective of the ability to monitor or control that other party. | (Mayer et al., 2006) |
| Effects of personality traits on user trust in human–machine collaborations (Zhou et al., 2020) | The attitude that an agent will help achieve an individual's goals in a situation characterized by uncertainty and vulnerability. | (Lee & See, 2004) |
| Exploring trust of mobile applications based on user behaviors: an empirical study (Yan et al., 2013) | His/her belief on whether the application could fulfill a task as expected (the trustworthiness of mobile applications relates to their dependability, security, and usability). | Own definition and referenced (Avizienis et al., 2004) |
| Human trust in artificial intelligence: review of empirical research (Glikson & Woolley, 2020) | The willingness of a party to be vulnerable to the actions of another party based on the expectation that the other will perform a particular action important to the trustor, irrespective of the ability to monitor or control that other party | (Mayer et al., 2006) |
| Incorporating behavioral trust theory into system development for ubiquitous applications (Hoffmann & Söllner, 2014) | The belief that an agent will help achieve an individual's goal in a situation characterized by uncertainty and vulnerability. | (Lee & See, 2004) |
| Trust and distrust in information systems at the workplace (Thielsch et al., 2018) | The willingness to depend on and be vulnerable to an Information System in uncertain and risky environments. | (Gefen et al., 2008; Mayer et al., 2006; Meeßen et al., 2020; Wang & Emurian, 2005) |

### 3.2.3. User characteristics influencing user trust

User characteristics dominated the findings and were thus divided by user inherent characteristics ($N = 3$ articles), user acquired characteristics ($N = 4$ articles), user attitudes ($N = 10$ articles), and user external variables ($N = 6$ articles) (Table 5).

#### 3.2.3.1. User inherent characteristics (i.e., personality traits, gender, and self-trust). Zhou et al. (2020) found that user personality traits influenced user predictive decision making and trust in AI-enabled systems. The study used the big five personality traits (Gosling et al., 2003) and found that Low Openness traits (practical, conventional, prefers routine) had the highest trust, followed by Low Conscientiousness (impulsive, careless, disorganized), Low Extraversion (quiet, reserved, withdrawn), and High Neuroticism (anxious, unhappy, prone to negative emotions). Given that personality traits were found to influence user trust, a user interface was suggested to include modules to identify and inform user personality traits to users. This would allow users to be aware of how their personality traits influenced their decision-making when interacting with an AI-enabled system.

Additionally, women were found to be more likely to yield a higher level of trust in an AI-enabled system (Morana et al., 2020). Another article looking into user self-trust found that a user was likely to use their own skills to gather and analyze information to decide whether to trust a system (Duffy, 2017).

#### 3.2.3.2. User acquired characteristics (i.e., user experiences and educational levels). A user's previous experience with a provider or producer of an AI-enabled system was found to influence user trust (Foehr & Germelmann, 2020; Yan et al., 2013). Positive experiences with a system allowed the user

to be rooted deeply in the provider's or producer's ecosystem, enabling the transfer of such trust to other systems from the same provider or producer. Importantly, a user's need or dependency to use a specific AI-enabled system overruled previous negative experiences with the system, especially if the negative experiences were predictable (Yan et al., 2013).

Generally, users without a college education were less likely to trust an AI-enabled system than those with a college education (Elkins & Derrick, 2013). The study concluded that this finding might be rooted in the unfamiliarity with or the perception of non-benevolence of AI-systems. Nevertheless, the study also found that trust increased over time along with growing familiarity with the system, including when the initial trust level in the AI-enabled system was relatively low.

#### 3.2.3.3. User attitudes (i.e., user acceptance and readiness, needs and expectations, judgment and perceptions). User acceptance and readiness of an AI-enabled system were found to be key determinants of user trust (Foehr & Germelmann, 2020; Khosrowjerdi, 2016; Klumpp & Zijm, 2019; Smith, 2016). Two studies suggested that addressing challenges such as artificial divide (Klumpp & Zijm, 2019) and user uncertainties (Hoffmann & Söllner, 2014) were fundamental for promoting user acceptance and readiness. The first study defined the artificial divide as the ability or lack thereof to cooperate successfully with AI-enabled systems (Klumpp & Zijm, 2019, p. 6). The study outlined that users might be divided by their motivation (e.g., intention to use) and technical competence toward AI-enabled systems. Users with low motivation and low technical competence were the risk group and needed more attention to enable their acceptance and readiness to use AI-enabled systems. The study highlighted the importance of

**Table 5.** The 23 included articles and factors influencing user trust.

| Title (alphabetically) | Socio-ethical considerations (N = 8 articles) | Technical and design features (N = 12 articles) | Inherent characteristics (N = 3 articles) | Acquired characteristics (N = 3 articles) | Attitudes (N = 10 articles) | External variables (N = 5 articles) |
|---|---|---|---|---|---|---|
| | | | User characteristics (N = 22 articles) | | | |
| "Let me explain!": exploring the potential of virtual agents in explainable AI interaction design (Weitz et al., 2021). | | ✓ | | | | |
| A comparison of consumer perception of trust-triggering appearance features on Indian group buying websites (Sharma, 2015) | | ✓ | | | | |
| A qualitative research framework for the design of user-centered displays of explanations for machine learning model predictions in healthcare (Barda et al., 2020) | | ✓ | | | | |
| A review of theory-driven models of trust in the online health context (Khosrowjerdi, 2016) | | | | | ✓ | |
| Alexa, can I trust you? Exploring consumer paths to trust in smart voice-interaction technologies (Foehr & Germelmann, 2020) | ✓ | ✓ | | ✓ | ✓ | ✓ |
| An extended framework for recovering from trust breakdowns in online community settings (Binmad et al., 2017) | ✓ | | | | | |
| Brokerbot: a cryptocurrency chatbot in the social-technical gap of trust (Lee et al., 2021) | ✓ | | | | ✓ | |
| Building e-commerce satisfaction and boosting sales: the role of social commerce trust and its antecedents (Lin et al., 2019) | ✓ | | | | | ✓ |
| Effects of personality traits on user trust in human–machine collaborations (Zhou et al., 2020) | | | ✓ | | | ✓ |
| Exploring trust of mobile applications based on user behaviors: an empirical study (Yan et al., 2013) | | | | ✓ | | ✓ |
| Human trust in artificial intelligence: review of empirical research (Glikson & Woolley, 2020) | | ✓ | | | | |
| Incorporating behavioral trust theory into system development for ubiquitous applications (Hoffmann & Söllner, 2014) | | | | | ✓ | |
| Legal, regulatory, and ethical frameworks for development of standards in artificial intelligence (AI) and autonomous robotic surgery (O'Sullivan et al., 2019) | ✓ | ✓ | | | | |
| Logistics innovation and social sustainability: how to prevent an artificial divide in human-computer interaction (Klumpp & Zijm, 2019) | | | | | ✓ | |
| Online trust and health information websites (Corritore et al., 2012) | | | | | ✓ | |

**Table 5.** Continued.

| | User trust (*N* = 23 articles) | | | | | |
|---|---|---|---|---|---|---|
| | | | User characteristics (*N* = 22 articles) | | | |
| Title (alphabetically) | Socio-ethical considerations (*N* = 8 articles) | Technical and design features (*N* = 12 articles) | Inherent characteristics (*N* = 3 articles) | Acquired characteristics (*N* = 3 articles) | Attitudes (*N* = 10 articles) | External variables (*N* = 5 articles) |
| Privacy and trust attitudes in the intent to volunteer for data-tracking research (Smith, 2016) | | | | | ✓ | |
| The automation of leadership functions: would people trust decision algorithms? (Höddinghaus et al., 2021) | ✓ | ✓ | | | ✓ | |
| The effect of anthropomorphism on investment decision-making with robo-advisor chatbots (Morana et al., 2020) | | ✓ | ✓ | | | |
| The interplay between emotional intelligence, trust, and gender in human-robot interaction: a vignette-based study (Law et al., 2021) | | ✓ | | | | |
| The sound of trust: voice as a measurement of trust during interactions with embodied conversational agents (Elkins & Derrick, 2013) | ✓ | ✓ | | ✓ | | |
| Effect of AI explanations on human perceptions of patient-facing AI-powered healthcare systems (Zhang, Genc, et al., 2021) | | ✓ | | | ✓ | |
| Trusting me, trusting you: evaluating three forms of trust on an information-rich consumer review website (Duffy, 2017) | | | ✓ | | | ✓ |
| Trust and distrust in information systems at the workplace (Thielsch et al., 2018) | ✓ | ✓ | | | ✓ | |

analyzing artificial divide elements (e.g., rejection of an AI-enabled system) and addressing challenges properly (e.g., early stage user involvement, training, enhanced user experience and empowerment) to foster user trust and prevent mistrust (Klumpp & Zijm, 2019). The second study suggested that user uncertainties had to be addressed by identifying and prioritizing the uncertainties and their antecedents in relation to a specific AI-enabled system, improving user understandability, sense of control, and information accuracy (Hoffmann & Söllner, 2014). A decrease in user uncertainties was suggested as an increase in user trust towards the specific AI-enabled system. Ironically, another study found that perceived imposition or inescapability of AI-enabled systems in general, i.e., a belief that AI-enabled systems would be a part of human daily life nonetheless, would initiate user trust as users perceived that trusting the systems was the only option (Foehr & Germelmann, 2020).

User needs and expectations of AI-enabled systems included user intention to use an AI-enabled system (Khosrowjerdi, 2016), relevance of technical system quality (e.g., reliability) and information quality (e.g., credibility) to users (Thielsch et al., 2018), as well as usefulness of an AI-enabled system to its users (Foehr & Germelmann, 2020). In general, user expectations of an AI-enabled system might not be aligned with the intention of the system's investors and developers (Lee et al., 2021). This might result in the system being operated in a way that was unforeseen by investors or developers, hitting and missing the target user expectations. The mismatch between user expectations and experiences was suggested to be a risk to user trust and needed to be addressed, especially when users were heavily dependent on specific AI-enabled systems (Lee et al., 2021; Thielsch et al., 2018).

For user judgement and perceptions, the key elements found to be affecting user trust in an AI-enabled system included perceived credibility (e.g., expertise, honesty, reputation, and predictability), risk (i.e., likelihood and severity of negative outcomes), and ease of use (e.g., searching, transacting and navigating) (Corritore et al., 2012; Foehr & Germelmann, 2020) as well as perceived benevolence, integrity and transparency (Elkins & Derrick, 2013; Höddinghaus et al., 2021). Importantly, it was found that the relatability a user felt to an AI-enabled system determined the user's trust in the system (Thielsch et al., 2018; Zhang, Genc, et al., 2021). If user trust was to be fostered, the studies suggested that a focus was needed to increase user relatability to and understandability of an AI-enabled system's rationale and performance.

#### 3.2.3.4. User external variables (i.e., initial interactions, user interactions, cognitive load levels, time and usage).

When an AI-enabled system was introduced to a potential user through the user's close relatives, friends or partner, the potential user typically used this opportunity to collect information regarding the system's benevolence, ability, and integrity (Foehr & Germelmann, 2020). In this study, the potential users were aware that they were introduced to an AI-enabled system by their close relatives, friends or partner (Foehr & Germelmann, 2020). Importantly, initial trust was likely to be fostered as well. In review-based recommender systems, the quality of user interactions on an AI-enabled system's platform was found to be a determinant of user trust (Duffy, 2017; Lin et al., 2019). For example, perceived similarities between users (e.g., preferences and interests) were taken into consideration when evaluating others' reviews (Duffy, 2017; Lin et al., 2019). Creating an effective environment where users were willing to exchange social support and share high-quality reviews was suggested as crucial to foster and maintain user trust (Lin et al., 2019). Another important determinant of user trust was the user's cognitive load when interacting with an AI-enabled system (Zhou et al., 2020). When under a low cognitive load, the user was more willing to trust a system enabled by a greater availability of the user's cognitive resources which allowed more confidence and willingness to analyze and understand the AI-enabled system.

One study found that user trust increased as more time was spent interacting with an AI-enabled system (Elkins & Derrick, 2013), likely as a result of understanding the system better and thus perceiving it had greater integrity (Elkins & Derrick, 2013; Lee et al., 2021). The study used an Embodied Conversational Agent (ECA) to ask participants 4 blocks of four questions, a total of 16 questions, similar to those screening questions asked at airports. Participants then were asked to rate their perceived trust of the ECA interviewer after each block of questions. After quantifying the trust ratings, the findings showed that user trust increased after each block of questions, regardless of the initial trust rating. User trust was thus suggested as multidimensional and continuous, and that human-system interactions were crucial to user trust development (Elkins & Derrick, 2013). Finally, usage was suggested as a reliable predictor of user trust; the more a user used an AI-enabled system, the more they trusted the system (Yan et al., 2013).

### 3.3. RQ3: How is user trust in AI-enabled systems measured?

A total of 16 studies (69.56%) used a survey either alone or in a combination with an interview or a focus group to measure user trust (Table 6). Of the 16 studies, 12 (75%) developed their own questionnaires, two (12.5%) developed their own as well as used previously developed questionnaires, and two (12.5%) used previously developed questionnaires. One of the 12 studies that developed their own questionnaires used it as a pre-survey to collect participant demographic characteristics and one study developed a questionnaire to collect participant preferred design options. Qualitative methods (e.g., interviews and/or focus groups) were the second most common methods used to measure user trust in six studies (26.09%), either as a stand alone or in combination with another method.

Twenty articles included participants in their studies, in which the number of participants ranged from 21 to 3423 participants ($M = 326.80$). Fourteen articles reported gender of the participants (range: 23.08–61.86% females; $M = 45.62\%$ females), and two articles reported 4% (Zhang, Genc, et al., 2021), 0.51% and 0.71% (Law et al., 2021) of gender claimed as other than male or female.

## 4. Discussion

This systematic literature review has identified 23 empirical studies which investigate how user trust is defined, factors influencing user trust, and methods for measuring user trust in AI-enabled systems. This section will discuss each research question separately.

### 4.1. RQ1: How is user trust in AI-enabled systems defined?

Of 23 studies, only seven explicitly define trust, while eight conceptualize it and the remaining nine provide neither. This is likely due to trust being an abstract concept that can be relatively difficult to define or generalize (Gebru et al., 2022; Gulati et al., 2019; Sousa et al., 2016), with dynamic characteristics, meaning that trust can change over time and in different contexts and situations (Elkins & Derrick, 2013). The difficulty in defining trust is reflected by findings that only one of the 23 included studies develop their own trust definition (Yan et al., 2013), whereas six studies use Mayer's and Lee and See's trust definitions (Lee & See, 2004; Mayer et al., 2006).

This finding confirms that there is more than one way to describe trust. Nevertheless, we propose that instead of pursuing better trust definitions or comparing which definitions are better, it is probably more beneficial to select the most appropriate trust definition according to the context, for example, based on the level of risk an output may affect a user. Mayer's trust definition, for example, may be able to provide a more accurate depiction of user trust in an AI-enabled system in which the output can have a significant personal impact to the user (e.g., personal finance, health). Whereas Lee and See's may be more accurate to be used for outputs that have a less personal impact to the user (e.g., complex information systems at workplace). The more accurate a trust definition is being used in specific contexts, the easier it is to understand user trust and factors influencing it.

### 4.2. RQ2: What factors influence user trust in AI-enabled systems

The first key finding is that user characteristics dominate the findings, reinforcing the importance of continuous

**Table 6.** The 23 included articles' titles, methods, participants/datasets, and questionnaires.

| Article title | The study main methodology | How the study measured user trust | The number of participants / datasets | The questionnaire(s) used |
|---|---|---|---|---|
| "Let me explain!": exploring the potential of virtual agents in explainable AI interaction design (Weitz et al., 2021) | An experimental user study | Survey | 60 participants (25 % females) | • Trust in Automation (TiA) questionnaire (Jian et al., 2000) • Developed a combination of 7-point Likert scales and open form questionnaires to collect qualitative and quantitative user feedback |
| A comparison of consumer perception of trust-triggering appearance features on Indian group buying websites (Sharma, 2015) | Survey | Survey | 110 students (35.5% females, mean age of 27.8 years old) | • The trust scale of Schlosser et al.'s (Schlosser et al., 2006) • Developed an additional participant assessment of the ten trust triggers on a Likert scale |
| A qualitative research framework for the design of user-centered displays of explanations for machine learning model predictions in healthcare (Barda et al., 2020) | Proposed conceptual framework | Focus group and a questionnaire to indicate preferred design options. | 21 pediatric critical care providers of differing clinical expertise (e.g., nurses, residents, fellows, attending physicians) | Developed a questionnaire to indicate participant preferred design options |
| A review of theory-driven models of trust in the online health context (Khosrowjerdi, 2016) | A literature review | A literature review | 12 theory-driven models | |
| Alexa, Can I Trust You? Exploring Consumer Paths to Trust in Smart Voice-Interaction Technologies (Foehr & Germelmann, 2020) | | A review of online comments and interviews | 600 generated comments, 26 users of smart speaker interviewees (23.08% females) | |
| An Extended Framework for Recovering From Trust Breakdowns in Online Community Settings (Bírnmad et al., 2017) | | Simulation experiment | At the outset, 50 service providers and 100 consumers | |
| Brokerbot: A Cryptocurrency Chatbot in the Social-technical Gap of Trust (Lee et al., 2021) | | Qualitative (i.e., interviews, group observations, focus group), usability testing | 8 novices (students), 2 developers, 5 investors, 50 cryptocurrency enthusiasts | |
| Building E-Commerce Satisfaction and Boosting Sales: The Role of Social Commerce Trust and Its Antecedents (Lin et al., 2019) | Developed a theoretical model | Survey | 903 (female 53.71%) | Developed a 32-question questionnaire |
| Effect of AI Explanations on Human Perceptions of Patient-Facing AI-Powered Healthcare Systems (Zhang, Genc, et al., 2021) | | Survey | 3423 Amazon Mechanical Turk (MTurk) employees (female 55%, other 4%) | Developed a post-study survey based on prior work (Ehsan et al., 2019) |
| Effects of personality traits on user trust in human–machine collaborations (Zhou et al., 2020) | Simulation experiment and case study | Survey | 42 participants (23.81% females, mean age of 30.4±8.5 years old) | • Developed a 9-point Likert scale to rate participant trust in the AI-enabled system's recommendations • Developed a 9-point Likert scale to rate participant cognitive load • Ten-Item Personality Inventory (TIPI) (Gosling et al., 2003) |
| Exploring trust of mobile applications based on user behaviors: an empirical study (Yan et al., 2013) | A psychometric method to examine the study hypothesis | Survey | 1120 university students (59.9% females) | Developed a 7-point Likert scale questionnaire priorly tested and optimized (Yan et al., 2008) |
| Human Trust in Artificial Intelligence: Review of Empirical Research (Glikson & Woolley, 2020) | | A literature review | 150 empirical research papers | |

**Table 6.** Continued.

| Article title | The study main methodology | How the study measured user trust | The number of participants / datasets | The questionnaire(s) used |
|---|---|---|---|---|
| Incorporating behavioral trust theory into system development for ubiquitous applications (Hoffmann & Söllner, 2014) | Proposed a process for software application development and a case study | Survey | 143 participants | Developed a 7-point Likert scale questionnaire |
| Legal, regulatory, and ethical frameworks for development of standards in artificial intelligence (AI) and autonomous robotic surgery (O'Sullivan et al., 2019) | | A literature review | | |
| Logistics Innovation and Social Sustainability: How to Prevent an Artificial Divide in Human-Computer Interaction (Klumpp & Zijm, 2019) | A theoretical framework and a case study | Expert workshop and interviews | An expert workshop with 10 employees from different disciplines (e.g. logistics, computer science) Semistructured interviews with 12 logistics management experts | |
| Online Trust and Health Information Websites (Corritore et al., 2012) | Proposed a model | Survey | 176 participants (55.9% females, mean age of 22.1 years old | A 34-item questionnaire previously developed (Corritore et al., 2007) |
| Privacy and trust attitudes in the intent to volunteer for data-tracking research (Smith, 2016) | | Survey | 110 students | Developed a questionnaire |
| The automation of leadership functions: Would people trust decision algorithms? (Höddinghaus et al., 2021) | | Survey | 333 workers (61.86% females, mean age of 48.55 years old | Developed a 60-question questionnaire |
| The effect of anthropomorphism on investment decision-making with robo-advisor chatbots (Morana et al., 2020) | | Survey | 183 participants (48.63% females, mean age of 23.77 ± 4.37 years old) | Developed a 24-question questionnaire |
| The Interplay Between Emotional Intelligence, Trust, and Gender in Human–Robot Interaction: A Vignette-Based Study (Law et al., 2021) | An experimental design | Survey | 198 participants through Amazon Mechanical Turk (47.98% females, 0.51% other, mean age of 34.96 ± 11.47). 421 usable data points (38.48% female, 0.71% other, mean age of 36.52 ± 11.85 years old). | A 24-item (Caruso & Salovey, 2004), a 4- item questionnaires (Mayer & Davis, 1999), and a 20-item of the Multidimensional Measure of Trust (MDMT) (Ullman & Malle, 2018) |
| The Sound of Trust: Voice as a Measurement of Trust During Interactions with Embodied Conversational Agents (Elkins & Derrick, 2013) | An experimental design | Survey | 88 participants (35% females, mean age of 25.45 ± 8.44 years old) | Developed a pre-survey for participant basic demographic information |
| Trust and distrust in information systems at the workplace (Thielsch et al., 2018) | | Survey and interviews | 30 professional interviewees (46.67% females, mean age of 32.03 ± 9.80 years old) 179 survey participants (53.63% females, mean age of 48.30 ± 9.22 years old) | Developed a questionnaire |
| Trusting me, trusting you: Evaluating three forms of trust on an information-rich consumer review website (Duffy, 2017) | | Survey and interviews | 30 the website user interviewees (53.33% females, mean age of 30 years old) 237 the website user survey participants (58% females) | Developed an 18-question questionnaire |

user involvement from system development through to the implementation and monitoring of AI-enabled systems (Khosrowjerdi, 2016; Klumpp & Zijm, 2019). The second key finding is that user trust can increase over time due to more user-system interactions (Elkins & Derrick, 2013; Lee et al., 2021), suggesting that low initial user trust is not fixed and can be improved. This finding highlights the importance of the user-AI interactions as a factor to foster user trust over time (Glikson & Woolley, 2020). It is likely that the interactions allow users to adjust their expectations and familiarize themselves to the system, resulting in increased trust in the system (Lee et al., 2021). For example, one included study that investigates how stakeholders of a cryptocurrency chatbot experience trust through the bot, found that the participants (users and developers) would start trusting the bot after initial interactions with the bot as part of "a journey" to discover the bot's features (Lee et al., 2021).

The third key finding is that different factors influence user trust based on different contexts and different characteristics of the users and systems. This highlights the importance of selecting and tailoring features of the system according to the targeted user group's characteristics and attributes. For example, technical and design features found to influence user trust can guide AI-enabled system design strategy (Weitz et al., 2021), as well as determining which technical and design features should be emphasized according to contexts and goals of the system tasks (Rheu et al., 2021). User characteristics evident to influence user trust can be used to optimize which AI-enabled systems, or their features, fit best for specific types of users. For example, user inherent characteristics can be used as a basis to determine which AI-enabled systems are the best fit and to adjust and improve the system design according to the inherent characteristics of a target group (Duffy, 2017; Morana et al., 2020; Zhou et al., 2020). Whereas, interventions such as user empowerment to foster, maintain or regain user trust in AI-enabled systems can be targeted specifically to improve user attitudes, experiences and external variables (i.e., factors that are more dynamic and open for change than those of user inherent characteristics) (Smith, 2019).

The fourth key finding is that socio-ethical considerations can pave the way in making sure that the environment where user-AI interactions happen is sufficiently conducive for these interactions to develop into trusted relationships (O'Sullivan et al., 2019). For example, the importance of explanations of AI-enabled systems have been consistently mentioned (Barda et al., 2020; Glikson & Woolley, 2020; O'Sullivan et al., 2019; Zhang, Bengio, et al., 2021), highlighting their potential role to improve user experiences and attitudes. Additionally, the findings show that there is still a significant challenge in setting up ethical-legal boundaries for usage of AI-enabled systems due to, for example, a gap between regulations and practices. In this case, a collective, multidisciplinary effort to close this gap is urgently needed (Hagendorff, 2020; Shneiderman, 2020c).

## 4.3. RQ3: How user trust in AI-enabled systems is measured

Over two-thirds (69.56%) of the included studies developed and used their own questionnaires to measure user trust (Table 6), illustrating that surveys are found to be the most common method to measure user trust. Qualitative methods (e.g., interviews or focus groups) were the second most used methods for measuring trust. Although qualitative methods are the most appropriate method to explore complex topics such as trust (Barda et al., 2020; Klumpp & Zijm, 2019), the caveat is that results are harder to compare than those of quantitative methods (e.g., surveys) and susceptible to varied interpretations. Nevertheless, these findings highlight different tools to measure user trust and are stated as a concern (Glikson & Woolley, 2020; Gulati et al., 2019). The concern relates to the fact that if user trust can be understood and measured in different ways, then being able to build upon the concept becomes challenging. A validated tool that allows empirical measurement of user trust across environments and contexts (Schepman & Rodway, 2022), such as the 12-item Human-Computer Trust Scale (HCTS) by Gulati et al. (2019), may be used to address this concern. Without doubt, more studies measuring and thus providing a more complete picture of user trust in AI-enabled systems are needed.

### 4.4. Limitations

This study has several limitations. First, the 23 identified studies have different contexts and some are rather specific. As such, generalization of findings may not be feasible and should be done with caution. Second, it is likely that there are trust definitions other than the ones identified from the 23 included studies. Third, our search terms may be perceived as too narrow and may have resulted in other potentially relevant studies being excluded. Nevertheless, the search terms still capture various types of AI-enabled systems used in different disciplines where the user-AI relationship is the focus. Fourth, our review does not include grey literature, which may have resulted in the exclusion of other potentially relevant work.

## 5. Conclusions and future aims

The user trust definitions, influencing factors, and measurement methods are crucial topics to be further explored as user trust in AI-enabled systems has been increasingly recognized and proven as a key element to foster adoption (Jacovi et al., 2021; Shin, 2021). Future studies should investigate and evaluate trust concepts and their applications in specific contexts with AI-enabled systems. Although several factors are found to influence user trust, it is still unclear how these factors fit to distinct types of users and contexts and/or change over time. The factor of time needs to be specifically investigated further and in greater detail to understand how influencing factors play a role at different points of time, and how user-AI interactions evolve into user-AI relationships. Surveys, interviews and focus groups, the most common methods found to measure user trust, are dependent on user perceptions that can be argued as being relatively subjective. Future research may explore other methods, possibly in addition to quantitative or qualitative methods, such as using psychophysiological signals (Barda et al., 2020; Gebru et al., 2022; Klumpp & Zijm, 2019), to gather a more objective insight towards understanding user trust in AI-enabled systems. Approximately half of the studies included in this review were conducted in the USA and Germany, highlighting that future research should be conducted in more diverse geographical locations to help understand how cultural factors influence user trust in AI-enabled systems (Jobin et al., 2019; Mohapatra & Kumar, 2019; Rheu et al., 2021).

Our findings highlight that fostering user trust in AI-enabled systems requires involvement from a multidisciplinary team from the early concept and ideation phases, as has also been suggested by other similar studies (Mohapatra, 2021; Panda & Mohapatra, 2021). Such a team should involve not only AI-enabled system developers, designers and target users, but also individuals with ethics, legal, behavioral, social sciences, and domain/industry expertise (Dwivedi et al., 2021). Integrating target user characteristics into technical and design development of AI-enabled systems needs to consider which characteristics are inherent, acquired, attitudes and external variables. This is because inherent user characteristics, for example, are less likely to change compared to user attitudes, thus requiring the systems to be adjusted accordingly. In contrast, user attitudes and external variables, for example, are probably more responsive to efforts to improve user experience (Zhang, Genc, et al., 2021). Ensuring quality interactions between users and AI-enabled systems requires adjusting the environment where these interactions happen by, for example, setting up mechanisms to foster, maintain and recover user trust. Importantly, efforts to calibrate the user-AI relationship requires finding the optimal balance that works for not only the user but also the system (DiSalvo et al., 2002; Fink, 2012; Gebru et al., 2022; Shneiderman, 2020b). This is because "we still believe that robots – as well as humans – need to be authentic in the way they are, to be 'successful' in a variety of dimensions." (Fink, 2012, p. 205).

## Author contributions

SS developed the concept and idea. All authors performed data collection, analysis, and interpretation. TAB and AK quality checked the results and interpretation for consistency and led the writing of the manuscript. All authors contributed to the writing and agreed with the final version of the manuscript.

## Disclosure statement

No potential conflict of interest was reported by the author(s).

## Funding

This work was supported by AI-Mind that has received funding from the European Union's Horizon 2020 research and innovation programme under grant agreement No 964220. This study was partly funded by the Trust and Influence Programme [FA8655-22-1-7051], European Office of Aerospace Research and Development, and US Air Force Office of Scientific Research.

## ORCID

Sonia Sousa  http://orcid.org/0000-0002-5865-1389

## About the authors

**Tita Alissa Bach** holds a doctorate degree in behavioural and social sciences from Groningen University, the Netherlands, with a strong focus in Human Factors and Organisational Psychology. Her research focuses on applying Human Factors principles in an environment or organization, especially in the context where technology is being used.

**Amna Khan** is a Junior Researcher and PhD student in Information Society Technologies at Tallinn University in Estonia. Her research focuses on the creation of a framework for an adaptive system to enhance knowledge reuse to develop trustworthy and transparent systems. She earned a master's degree in Mechatronics Engineering (2017).

**Harry Hallock** holds his doctorate degree in Cognitive Neurosciences from the University of Sydney, Australia. He has experience in product development and large-scale project management. His current work focuses on understanding trustworthy AI in healthcare, federated networks for health data, governance, management and quality assurance of health data.

**Gabriela Beltrão** candidate in Information Society Technologies at Tallinn University, Estonia. Her research targets trust in technology from a human-centered perspective, focusing on differences in trust across cultures and their implications for the design.

**Sonia Sousa** is an Associate Professor of Interaction Design, Tallinn University, Estonia, in HCI, and the head of the Join Online MSc in Interaction Design. Her funded projects include NGI-Trust, CHIST-ERA, Horizon 2020, and AFOSR. She was nominated for AcademiaNet - the expert database for outstanding female academics.